\documentclass[fleqn,12pt,twoside]{article}
\usepackage{espcrc1}

\usepackage{graphicx}

\newcommand{\ket}[1]{\left|#1\right\rangle}
\newcommand{\bra}[1]{\left\langle#1\right|}
\def\Hcontr{{\cal H}_{\rm ctrl}}
\def\P{{\cal P}}

\title{Noise and Decoherence in Quantum Two-Level Systems}

\author{Alexander Shnirman\address[TFP]{Institut f\"ur 
                     Theoretische Festk\"orperphysik, Universit\"at Karlsruhe,  
                     D-76128 Karlsruhe, Germany},
        Yuriy Makhlin\addressmark[TFP]%
                     \address[Landau]{Landau Institute for Theoretical 
                     Physics, Kosygin St. 2, 117940, Moscow, Russia},
        and
        Gerd Sch\"on\addressmark[TFP]\address[FZK]
                     {Forschungszentrum Karlsruhe, 
                      Institut f\"ur Nanotechnologie, 
                      D-76021 Karlsruhe} 
        }

\begin{document}

\maketitle

\begin{abstract}

Motivated by recent experiments with Josephson-junction circuits we 
reconsider decoherence effects in quantum two-level systems (TLS).  
On one hand, the experiments demonstrate the importance of $1/f$ noise, 
on the other hand, by operating at symmetry points one can suppress noise
effects in linear order. We, therefore, analyze noise sources with
a variety of power spectra, with linear or quadratic coupling, which are
longitudinal or transverse relative to the eigenbasis of the
unperturbed Hamiltonian.
To evaluate the dephasing time for transverse $1/f$ noise second-order
contributions have to be taken into account.
Manipulations of the quantum state of the
TLS define characteristic time scales.  We discuss the consequences 
for relaxation and dephasing processes.
\end{abstract}

\section{Introduction}

The dynamics of quantum two-level systems has always been at the focus
of interest, but recently has attracted increased attention because of the
ideas of quantum computing.  Several systems have been suggested as
physical realizations of quantum bits allowing for the needed controlled 
manipulations,
and for some of them first elementary steps have been demonstrated in
experiments (see reviews in Ref. \cite{Realizations}).  A crucial
requirement is the preservation of phase coherence in the presence of
a noisy environment.  In solid-state realizations, incl.\ those based
on superconducting circuits, major noise sources are the electronic
control circuit as well as material-specific fluctuations, e.g., in
the substrate.  In many cases one lacks a detailed microscopic
description of the noise source, but frequently it is sufficient to model
the environment by a bath of harmonic oscillators with frequency
spectrum adjusted to reproduce the observed power spectrum.  The
resulting ``spin-boson models'' have been studied in the literature
(see the reviews \cite{LeggettRMP,WeissBook}), in particular the one
with linear coupling and ``Ohmic'' spectrum.  On the other hand,
several recent experiments appear to be
described by spin-boson models with
more general couplings and different power spectra.  In this article
we, therefore, describe these extensions of the spin-boson model and analyze
how more general noise sources  and coupling schemes affect relaxation and 
dephasing processes.

In the following Section we summarize what is known about
relaxation and dephasing processes for spin-boson models with linear
coupling. We add several extensions, including a discussion how these
processes are induced by specific quantum manipulations. We also
present some exact results, shedding light on the important question
of low-temperature dephasing.  This analysis, as well as the rest of 
this article are applicable to any quantum two-state system subject to
noise with appropriate coupling and power spectrum.  However, since
the extensions have been motivated by recent experiments with
Josephson-junction qubits, we shortly present in Section 3 the
physics of these systems. We indicate how, by proper choice of
the operating point, one can tune to zero the linear coupling to the
environment. Section 4 covers the extensions of the spin-boson model,
which were motivated by the recent experiments. This includes noise
sources with 
a variety of power spectra, with linear or quadratic coupling, which are
longitudinal or transverse relative to the eigenbasis of the
unperturbed Hamiltonian. For transverse $1/f$ noise we evaluate also 
second-order
contributions, which turn out to be comparable to the lowest order.
We conclude with a summary.
The technical parts 
of our work, which are based on an analysis of the time evolution of 
the reduced density matrix of the quantum two-state system will be 
described in a separate publication.

\section{Spin-boson model}
\label{Sec:spin-boson}

In this section we review the theory and properties of the spin-boson model,
which have been studied extensively before (see the reviews
\cite{LeggettRMP,WeissBook}).  A quantum two-level system coupled to
an environment is modeled by a spin degree of freedom in a
magnetic field coupled linearly to an oscillator bath with Hamiltonian
\begin{equation}
{\cal H}=\Hcontr + \sigma_z \sum_j c_j (a^{\phantom
 \dagger}_j+a^\dagger_j) + {\cal H}_{\rm b} \; .
\label{Eq:SpinBoson}
\end{equation}
The controlled part is
\begin{eqnarray}
\label{Hcontr}
\Hcontr = -\frac{1}{2}B_z\;\sigma_z- \frac{1}{2}B_x\;\sigma_x =
-\frac{1}{2} \Delta E \,(\cos\theta\;\sigma_z+\sin\theta\;\sigma_x) \; ,
\end{eqnarray}
while the oscillator bath is described by
\begin{equation}
{\cal H}_{\rm b}= \sum_j \hbar \omega_j \, a^\dagger_ja^{\phantom
\dagger}_j \; .
\label{Eq:FreeBosons}
\end{equation}
In the second form of Eq.~(\ref{Hcontr}) we introduced the mixing angle
$\theta\equiv \tan^{-1} (B_x/B_z)$, which depends on the direction of
the magnetic field in the $x$-$z$-plane, and the energy splitting
between the eigenstates, $\Delta E = \sqrt{B_x^2+B_z^2}$. In the
standard spin-boson model it is assumed that the bath ``force'' operator
$X=\sum_j c_j (a^{\phantom \dagger}_j+a^\dagger_j)$ couples linearly
to $\sigma_z$.

In thermal equilibrium the Fourier transform of the symmetrized
correlation function of this operator is given by
\begin{equation}
\label{Eq:X-J}
S_X(\omega)
\equiv \left\langle [ X(t), X(t') ]_+ \right\rangle_\omega
= 2 \hbar J(\omega) \coth \frac{\hbar \, \omega}{2 k_{\rm B}T} \;.
\end{equation}
Here the bath spectral density has been introduced, defined by
\begin{equation}
J(\omega) \equiv {\pi\over\hbar}\sum_j c_j^2 \;
\delta(\omega-\omega_j) \,.
\label{Eq:CL-spectrum}
\end{equation}
At low frequencies it typically has a power-law behavior. Of particular 
interest is the ``Ohmic
dissipation'', corresponding to a spectrum which is linear at low 
frequencies up to some high-frequency cutoff $\omega_c$,
\begin{equation} 
\label{Eq:Linear_Spectrum}
J(\omega)=\frac{\pi}{2}\;\alpha\,\hbar \omega \, 
\Theta(\omega_c - \omega) \; .
\end{equation} 
The dimensionless parameter $\alpha$ reflects the strength of
dissipation.  In a physical system it depends on the amplitude of 
the noise and the coupling strength.
Here we concentrate on weak damping, $\alpha \ll 1$,
since this limit is relevant for quantum-state engineering.
Still we distinguish two regimes: the Hamiltonian-dominated regime,
which is realized when $\Delta E \gg \alpha \, k_{\rm B} T$, and the
noise-dominated regime, which is realized, e.g., at degeneracy points
where $\Delta E \to 0$.

In the Hamiltonian-dominated regime, $\Delta E \gg \alpha \, k_{\rm B}
T$, it is natural to describe the evolution of the system in the
eigenbasis, $\ket{0}$ and $\ket{1}$, of $\Hcontr$:
\begin{eqnarray}
\label{Eigen_Basis}
        \ket{0} &=& \;\;\; \cos{\theta\over2} \ket{\uparrow} +
\sin{\theta\over2} \ket{\downarrow} \nonumber \\ \ket{1} &=& -
\sin{\theta\over2} \ket{\uparrow} + \cos{\theta\over2}
\ket{\downarrow} \ .
\end{eqnarray}
Denoting by $\tau_x$ and $\tau_z$ the Pauli matrices in the eigenbasis,
we have
\begin{equation}
\label{Eq:Spin_Boson_Eigen_Basis}
{\cal H} = -{1\over 2}\Delta E\, \tau_z + (\sin\theta\;\tau_x +
\cos\theta\;\tau_z) \; X + {\cal H_{\rm b}} \ .
\end{equation} 
\vspace{3mm}

\subsection{Relaxation and dephasing}

Two different time scales describe the evolution in the
spin-boson model~\cite{LeggettRMP,WeissBook,Weiss1,Weiss2}.  The first is the
dephasing time scale $\tau_\varphi$. It characterizes the decay of the
off-diagonal elements of the qubit's reduced density matrix
$\hat\rho(t)$ in the preferred eigenbasis (\ref{Eigen_Basis}), or,
equivalently  of the expectation values of the operators
$\tau_{\pm} \equiv (1/2) (\tau_x \pm i \tau_y)$. Dephasing processes lead to the 
following long-time dependence:
\begin{equation}
\label{Eq:Rho_pm_dephasing} 
\langle \tau_{\pm}(t) \rangle 
\equiv \mbox{tr}\,[\tau_{\pm} \hat\rho(t) ]
\propto \langle \tau_{\pm}(0) \rangle\; e^{\mp
i\Delta E t/\hbar}\; e^{-t/\tau_\varphi} \ .
\end{equation}
(Other time dependences will be discussed below). 
The second, the relaxation time scale $\tau_{\rm relax}$, characterizes how
the diagonal entries tend to their thermal equilibrium values:
\begin{equation}
\label{Eq:Rho_z_relaxation}
\langle \tau_z(t) \rangle - \tau_z(\infty) 
\propto
e^{-t/\tau_{\rm relax}} \ ,
\end{equation}
where $\tau_z(\infty)=\tanh(\Delta E/2k_{\rm B}T)$.

In Refs.~\cite{LeggettRMP,WeissBook} the dephasing and relaxation
times for the spin boson model with Ohmic spectrum were evaluated in a 
path-integral
technique.  In the regime $\alpha \ll 1$ it is easier to
employ the perturbative (diagrammatic) technique developed in 
Ref.~\cite{Schoeller_PRB} and the standard Bloch-Redfield approximation. 
The rates are~\footnote{Note that in the literature usually the evolution
of $\langle \sigma_z(t) \rangle$ has been studied.  To establish the
connection to the results (\ref{Eq:relaxation},\ref{Eq:dephasing}) one
has to substitute
Eqs.~(\ref{Eq:Rho_pm_dephasing},\ref{Eq:Rho_z_relaxation}) into the
identity $\sigma_z = \cos\theta\;\tau_z + \sin\theta\;\tau_x$.
Here we neglect renormalization effects, since they are weak
for $\alpha \ll 1$.}
\begin{eqnarray}
\Gamma_{\rm relax} \equiv \tau_{\rm relax}^{-1}
&= \pi\alpha\;\sin^2\theta \; 
\displaystyle \frac{\Delta E}{\hbar} 
\coth\displaystyle\frac{\Delta E}{2k_{\rm B}T}
&= 
\frac{1}{\hbar^2} \sin^2\theta \; S_X\left(\omega =
\Delta E/\hbar\right)
\label{Eq:relaxation}
\; , \\ 
\Gamma_\varphi \equiv \tau_\varphi^{-1}
&= \frac{1}{2}\;\Gamma_{\rm relax} +
\pi\alpha\;\cos^2\theta\;\displaystyle  \frac{2k_{\rm B}T}{\hbar} 
&=
\frac{1}{2}\;\Gamma_{\rm relax} + \frac{1}{\hbar^2} \cos^2\theta \, 
S_X(\omega = 0)
\label{Eq:dephasing}
\; .
\end{eqnarray}
We observe
that only the ``transverse'' $\tau_x$-component of the fluctuating
field, proportional to $\sin\theta$, induces transitions between the
eigenstates (\ref{Eigen_Basis}) of the unperturbed system.  Thus the
relaxation rate (\ref{Eq:relaxation}) is proportional to
$\sin^2\theta$.  The ``longitudinal'' $\tau_z$-component of the
fluctuating field, proportional to $\cos\theta$, does not induce
relaxation processes.  It does contribute, however, to dephasing since
it leads to fluctuations of the eigenenergies and, thus, to a
random relative phase between the two eigenstates.  This is the origin
of the contribution to the ``pure'' dephasing rate $\Gamma^*_\varphi$ 
(\ref{Eq:dephasing}), 
which is proportional to $\cos^2\theta$. Relaxation and pure dephasing
contribute to the rate $\Gamma_\varphi = \frac{1}{2} \Gamma_{\rm relax} + 
\Gamma^*_\varphi$.

The last forms of Eqs.~(\ref{Eq:relaxation}) and (\ref{Eq:dephasing}) 
express the two rates in terms of the noise power spectrum at the
relevant frequencies. These are the
level spacing of the two-state system and zero frequency, respectively. 
The expressions apply in the weak-coupling limit for 
spectra which are regular at these frequencies (see Section~\ref{Sec:1/f}
for $1/f$ noise). In general, fluctuations with frequencies 
in an interval of width $\Gamma_{\rm relax}$ around $\Delta E/\hbar$ 
and of width $\Gamma_\varphi$ around zero are involved.

The equilibration is due to two processes,
excitation $\ket{0}\to\ket{1}$ and relaxation 
$\ket{1}\to\ket{0}$, with rates
$\Gamma_{+/-} \propto
\langle X(t)X(t')\rangle_{\omega = \pm\Delta E/\hbar}$.
Both rates are related by a detailed balance condition, and
the equilibrium value $\tau_z (\infty )$ depends on both.
On the other hand, $\Gamma_{\rm relax}$ is determined by the 
sum of the two rates, i.e., the symmetrized noise 
power spectrum $S_X$. 

In the environment-dominated regime, $\Delta E \ll \alpha \, k_{\rm B}T$,
the coupling to the bath is the dominant part of the total Hamiltonian.
Therefore, it is more convenient to discuss the problem in the
eigenbasis of the observable $\sigma_z$ to which the bath is
coupled. The spin can tunnel incoherently between the two eigenstates
of $\sigma_z$.  One can again employ the perturbative analysis
\cite{Schoeller_PRB} but use directly the Markov instead of the
Bloch-Redfield approximation.  The resulting rates are given by
\begin{eqnarray}
\Gamma_{\rm relax}  &\approx& B_x^2 /(2\pi\hbar\alpha k_{\rm B}T) \
,\\
\nonumber
\Gamma_{\rm \varphi} &\approx& 2\pi\alpha k_{\rm B}T/\hbar \ .
\label{Eq:Tau_Deph_Zeno}
\end{eqnarray}  
In this regime the dephasing is much faster than the relaxation.  In
fact, we observe that as a function of temperature the dephasing and
relaxation rates evolve in opposite directions.  
The $\alpha$-dependence of the relaxation rate is an indication of the Zeno 
(watchdog) effect~\cite{Harris_Stodolsky}: the environment frequently
``observes'' the state of the spin, thus preventing it from tunneling.

\subsection{Sensitivity to the initial state}

To get further insight into dephasing phenomena we analyze 
some experimental scenarios. In particular, we discuss
their sensitivity to details of the preparation of the initial state (cf., 
e.g., Ref.~\cite{GPW}). Such a preparation is a necessary ingredient in an
experiment probing dephasing processes directly. We observe that
depending on the time scales of the 
preparation part of the bath oscillators (the fast ones) may follow
the two-level system adiabatically, merely renormalizing its
parameters, while others (the slow ones) lead to dephasing.
We illustrate these results by considering the exactly 
solvable limit $\theta = 0$.

Dephasing processes are contained in the time evolution of the quantity 
$\langle \tau_+(t)\rangle$ obtained after tracing out the bath. This quantity 
can be evaluated analytically for $\theta=0$ 
(in which case $\tau_+ = \sigma_+$), for an initial state described 
by a factorized 
density matrix $\hat\rho(t=0) = \hat\rho_{\rm spin} \otimes \hat\rho_{\rm 
bath}$ (i.e., the TLS and the bath are disentangled). 
We find
\begin{equation}
\langle \sigma_+(t) \rangle \equiv \P(t)\; e^{-i\Delta E t}\;
\langle \sigma_+(0) \rangle
\;,
\qquad
\mbox{where}
\qquad
\label{Eq:P(t) factorized}
\P(t)= {\rm Tr}(e^{i\Phi(0)/2} e^{- i\Phi(t)} e^{i\Phi(0)/2} \hat\rho_{\rm 
bath})
\,.
\end{equation}
Here the bath operator $\Phi$ is defined as
\begin{equation}
\Phi \equiv i \sum_j \frac{2c_j}{\hbar\omega_j}
(a^{\dagger}_j-a^{\phantom \dagger}_j)
\,,\end{equation}
and its time evolution
is determined by the bare bath Hamiltonian, 
$\Phi(t)=e^{i{\cal H}_{\rm b}t} \Phi e^{-i{\cal H}_{\rm b}t}$.
To derive Eq.~(\ref{Eq:P(t) factorized}) we made use of a unitary
transformation by $U \equiv \exp\left(-i \sigma_z \Phi/2 \right)$.
In the new basis the bath and the TLS decouple, which allows for the exact 
solution.

The expression~(\ref{Eq:P(t) factorized}) applies for any state of the bath 
(as long as it is factorized from the spin). In 
particular, we can assume that the spin was initially (for $t\le0$) kept in the 
state $\ket{\uparrow}$ and the bath has relaxed to the thermal equilibrium 
distribution for this spin value: $\hat\rho_{\rm bath} = \hat\rho_{\uparrow}
\equiv Z_{\uparrow}^{-1} e^{-\beta {\cal H}_{\uparrow}}$, where
${\cal H}_{\uparrow} = {\cal H}_{\rm b}+\sum_j c_j 
(a^{\phantom\dagger}_j+a^\dagger_j)$. 
In this case we can rewrite the density matrix as
$\hat\rho_{\rm bath} =e^{i\Phi/2}\hat\rho_{\rm b} e^{-i\Phi/2}$,
with the density matrix of the decoupled bath given by
$\hat\rho_{\rm b}\equiv Z_{\rm b}^{-1}e^{-\beta {\cal H}_{\rm b}}$, and obtain
\begin{equation}
\label{Eq:P(t)}
\P(t) = P(t) \equiv {\rm Tr}\,\left(e^{-i\Phi(t)}\,e^{i\Phi}\,\hat\rho_{\rm
b} \right) \; .
\end{equation}
The latter expression (with Fourier transform $P(E)$) has been studied
extensively in the  literature
\cite{LeggettRMP,P(E)_Panyukov_Zaikin,P(E)_Odintsov,P(E)_Nazarov,P(E)_Devoret}.
It can be expressed as $P(t)=\exp K(t)$, where
\begin{eqnarray}
\label{K(t)}
        K(t) = {4\over \pi\hbar} \int_0^{\infty} d\omega \,
        {J(\omega)\over\omega^2} \left[\coth\left({\hbar\omega\over2
        k_B T}\right)(\cos\omega t-1) -i\sin\omega t\right] \ .
\end{eqnarray}
For an Ohmic bath (\ref{Eq:Linear_Spectrum}) at nonzero temperature
and $t>\hbar/2k_{\rm B}T$, it reduces to
\begin{equation} 
\label{Eq:ReK_convergent_case}
{\rm Re} K(t) \approx -\frac{S_X(\omega=0)}{\hbar^2}t=
-\pi\,\alpha\,\frac{2 k_{\rm B}T}{\hbar}\,t \ .
\end{equation} 
Thus we reproduce Eq.~(\ref{Eq:Rho_pm_dephasing}) with the dephasing 
rate $\Gamma_\varphi$
given by (\ref{Eq:dephasing}) in the limit $\theta = 0$.

This result is easy to understand at the Golden rule level.
Pure dephasing processes are associated with transitions during 
which only the oscillators change their state, while the spin remains 
unchanged.  Since no energy is exchanged
only oscillators with  frequencies near zero contribute. 
An analysis of the Golden rule shows that the range is
the inverse of the relevant time scale, in the present case given by 
$\Gamma_\varphi$.
Since for an Ohmic model the spectral 
density $J(\omega)$ vanishes linearly as $\omega \rightarrow 0$,
exponential dephasing is found only at nonzero temperature  $T>0$. 
The situation is different in the presence of sub-Ohmic noise where, even 
at $T=0$, the low-frequency oscillators may cause exponential dephasing 
(see the discussion in Sec.~{\ref{Subsec:_1/f}).

For vanishing bath temperature, $T=0$, one still finds a decay
of $\langle \sigma_+(t)\rangle$, governed 
by ${\rm Re}\;K(t) \approx -2\;\alpha\;\ln(\omega_c t)$, which implies
a power-law decay 
\begin{equation}
\label{Eq:power_law_dephasing}
\langle\sigma_+(t)\rangle = (\omega_c t)^{-2\alpha} e^{- i\Delta E t/\hbar}
\langle\sigma_+(0)\rangle$ for $t > 1/\omega_c
\ .
\end{equation}
This result is beyond the Golden rule. In fact all oscillators up to 
the high-frequency cutoff $\omega_c$ contribute. 
While the power-law decay is a weak effect, it highlights the
influence of the initial state preparation  
and coherent manipulations on the dephasing. This will be particularly 
important in the sub-Ohmic and the super-Ohmic regime, as we will 
describe later.  

So far we discussed various initial states without specifying how they  
are prepared. We now consider some possibilities. A
(theoretical) one is to keep the spin and bath decoupled until 
$t=0$. In equilibrium  the initial state  which then enters
Eq.~(\ref{Eq:P(t) factorized}) is the product state
$\hat \rho_{\rm bath} = \hat \rho_{\rm b} \propto e^{-\beta {\cal H}_{\rm 
b}}$.
A more realistic possibility is to keep the bath 
coupled to the spin, while forcing the latter, e.g., by a strong external field,
to be in a fixed state, say $\ket{\uparrow}$. Then, at $t=0$, a 
sudden pulse of the external field is applied to change 
the spin state to a superposition, e.g., 
$\frac{1}{\sqrt{2}}(\ket{\uparrow} + \ket{\downarrow})$. If the bath has no
time to respond the resulting state at $T=0$ is
$\ket{\rm i} = \frac{1}{\sqrt{2}}
(\ket{\uparrow} + \ket{\downarrow}) \otimes \ket{{\rm
g}_{\uparrow}}$, where $\ket{{\rm g}_{\uparrow}}$ is the
ground state of ${\cal H}_{\uparrow}$.
Both components of this initial wave function now evolve in
time according to the Hamiltonian (\ref{Eq:SpinBoson}). 
The first, $\ket{\uparrow} \otimes \ket{{\rm
g}_{\uparrow}}$, which is an eigenstate of
(\ref{Eq:SpinBoson}), acquires only a trivial phase factor.  The time
evolution of the second component is more involved.  Up to a phase
factor it is given by $\ket{\downarrow} \otimes \exp(-i{\cal
H}_{\downarrow}t/\hbar) \ket{{\rm g}_{\uparrow}}$ where ${\cal
H}_{\downarrow}\equiv {\cal H}_{\rm b}-\sum_j c_j 
(a^{\phantom\dagger}_j+a^\dagger_j)$.  As the
state $\ket{{\rm g}_{\uparrow}}$ is not an eigenstate of 
${\cal H}_{\downarrow}$, entanglement between the spin and the
bath develops, and the coherence between the states of the spin is reduced by
the factor $|\,\bra{{\rm g}_{\uparrow}} \exp(-i{\cal
H}_{\downarrow}t/\hbar) \ket{{\rm g}_{\uparrow}}\,| < 1$.

In a real experiment of the type discussed the preparation pulse 
takes a finite time, during which the oscillators partially adjust to the 
changing spin state. For instance, the $(\pi/2)_x$-pulse 
which transforms the state $\ket{\uparrow}\to 
\frac{1}{\sqrt{2}} (\ket{\uparrow}+\ket{\downarrow})$, can be accomplished by
putting $B_z=0$ and $B_x=\Delta$ for a time span $\pi\hbar/2\Delta$.
In this case the oscillators 
with (high) frequencies, $\hbar \omega_j \gg \Delta$, 
follow the spin adiabatically. 
In contrast, the oscillators with low frequency, 
$\hbar \omega_j\ll\Delta$, do not change their state. 
Assuming that the oscillators can be split into these two 
groups, we see that just after the 
$(\pi/2)_x$-pulse the state of the system is 
$\frac{1}{\sqrt{2}}\left(\ket{\uparrow}\otimes\ket{{\rm
g}_{\uparrow}^{\rm h}} + \ket{\downarrow}\otimes\ket{{\rm
g}_{\downarrow}^{\rm h}}\right) \otimes \ket{{\rm g}_{\uparrow}^{\rm
l}}$ where the superscripts `h' and `l' refer to the high-
and low-frequency oscillators, respectively.
Thus, we arrive at a factorized initial state of the type as discussed 
above. However, only the low-frequency 
oscillators are factorized from the spin and contribute to dephasing during the 
subsequent evolution,
while the high-frequency modes give rise to renormalization effects.
Thus we reproduce the decay of $\langle \sigma_+(t)\rangle$ 
described above (i.e., for an Ohmic bath a power-law decay),
however, with a preparation-dependent cutoff frequency
$\hbar \omega_c = \Delta$. We may add that similar considerations 
apply for the analysis of the measurement process.

\subsection{Response functions}

In the limit $\theta = 0$ we can also calculate exactly the
linear response of $\tau_x=\sigma_x$ to a weak magnetic field in
the $x$-direction, $B_x(t)$:
\begin{equation}
\label{Eq:response function}
\chi_{xx}(t) = \frac{i}{\hbar}\;\Theta(t) \langle
\tau_x(t)\tau_x(0)-\tau_x(0)\tau_x(t)\rangle \ .
\end{equation}  
Using the equilibrium density matrix
\begin{equation}
\hat \rho^{\rm eq} = \frac{1}{1+e^{-\beta\Delta E}}
\left[\ket{\uparrow}\bra{\uparrow}\otimes \hat\rho_{\uparrow} +
e^{-\beta\Delta E} \ket{\downarrow}\bra{\downarrow}\otimes
\hat\rho_{\downarrow} \right] \ ,
\end{equation}
where $\hat\rho_{\uparrow} \propto \exp(-\beta {\cal H}_{\uparrow})$ is the
bath density matrix adjusted to the spin state $\ket{\uparrow}$ and
similar for $\hat\rho_{\downarrow}$, 
we obtain the susceptibility
\begin{equation}
\chi_{xx}(t) = -\frac{2\hbar^{-1}\Theta(t)}{1+e^{-\beta\Delta E}}\;{\rm Im}
\left[P(t)e^{-i\Delta E t} + e^{-\beta \Delta E}P(t)e^{i\Delta E
t}\right] \ .
\end{equation} 
The imaginary part of its Fourier transform is
\begin{equation}
\chi_{xx}''(\omega) = \frac{1}{2(1+e^{-\beta \Delta E})}
\left[P(\hbar\omega-\Delta E)+ e^{-\beta \Delta E} P(\hbar\omega+\Delta E)
\right] - ...(-\omega) \ .
\end{equation}
At $T=0$ and positive values of $\omega$ we use
the expression for $P(E)$ from Ref.~\cite{P(E)_Devoret}
to obtain
\begin{equation}
\chi_{xx}''(\omega) = \frac{1}{2}P(\hbar\omega-\Delta E)=
\Theta(\hbar\omega-\Delta E)
\frac{e^{-2\gamma\alpha}(\hbar\omega_c)^{-2\alpha}}{2\Gamma(2\alpha)}
(\hbar\omega-\Delta E)^{2\alpha -1} \ .
\end{equation}
We observe that the dissipative part
$\chi_{xx}''$ has a gap $\Delta E$, which corresponds to 
the minimal energy needed to flip the spin, 
and a power-law behavior  as $\omega$
approaches the threshold. This behavior of $\chi_{xx}''(\omega)$
parallels the {\it orthogonality catastrophe}
scenario~\cite{Mahan}. It implies that the ground state of the
oscillator bath for  different spin states,
$\ket{{\rm g}_{\uparrow}}$ and $\ket{{\rm g}_{\downarrow}}$, are
{\it macroscopically} orthogonal. In particular, for an Ohmic bath 
(\ref{Eq:Linear_Spectrum}) we recover the behavior typical for the problem of 
the X-ray absorption in metals~\cite{Mahan}.

The case of perpendicular noise, $\theta = \pi/2$, has been treated in
the literature (see e.g. Refs.~\cite{WeissBook,Keil_Schoeller_RG}).
 Then, in the 
coherent (Hamiltonian-dominated) regime, $\chi_{xx}''(\omega)$ 
shows a Lorentzian peak around the frequency corresponding 
to the (renormalized) level splitting $\Delta E$.
The width of the peak is determined by the dephasing associated with 
the relaxation processes,
i.e., by $\Gamma_\varphi = \Gamma_{\rm relax}/2$ as given in 
Eqs.~(\ref{Eq:relaxation}) and (\ref{Eq:dephasing}).

\section{Josephson junction qubits}

In this section we describe, by way of a specific example, some principles 
of quantum-state manipulations and the influence of various noise sources. 
For this purpose we consider a Josephson-junction qubit based on a 
superconducting single-charge box displayed
in Fig.~1a (for an extended discussion see Ref.~\cite{Our_RMP}). 
It consists of a small superconducting island coupled via
Josephson junctions with effective coupling energy $E_{\rm J}(\Phi_{\rm 
x})$ and the junction capacitance $C_{\rm J}$) to a superconducting
lead, and via a gate capacitor $C_{\rm g}$ 
to a gate voltage source $V_{\rm g}$ \cite{Saclay_Box}. 
The Hamiltonian of this system is
\begin{equation}
\label{Hbox}
{\cal H} = \frac{(Q-Q_{\rm g})^2}{2C} 
- E_{\rm J}(\Phi_{\rm x})\, \cos \varphi
\;\; \;\; \mbox{with} \;\;\; 
Q = \frac{\hbar}{i}\, \frac{\partial}{\partial(\hbar \varphi/2e)}.
\end{equation}
The excess Cooper-pair charge $Q=2ne$ on the island, which is an
integer multiple of $2e$,  and the phase difference
$\varphi$ across the Josephson junction are
conjugate variables \cite{Caldeira_Leggett_PRL81}, as indicated by the 
second part of Eq.~(\ref{Hbox}). The charging energy, with a characteristic 
scale $E_C \equiv e^2/2C$, depends on the total 
capacitance of the island, $C = C_{\rm J} + C_{\rm g}$, and is 
controlled by the  gate charge $Q_{\rm g}= 
C_{\rm g}V_{\rm g}$. We allow for a circuit with several
junctions such that the Josephson coupling energy can be controlled 
by an applied flux $\Phi_{\rm x}$. 

\begin{figure}[htb]
\parbox{10pc}{\includegraphics[width=10pc]{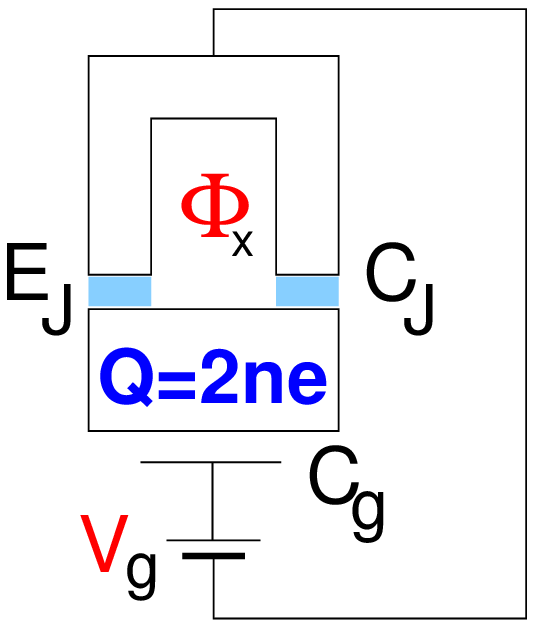}}
\hskip 1cm
\parbox{15pc}{\includegraphics[width=15pc]{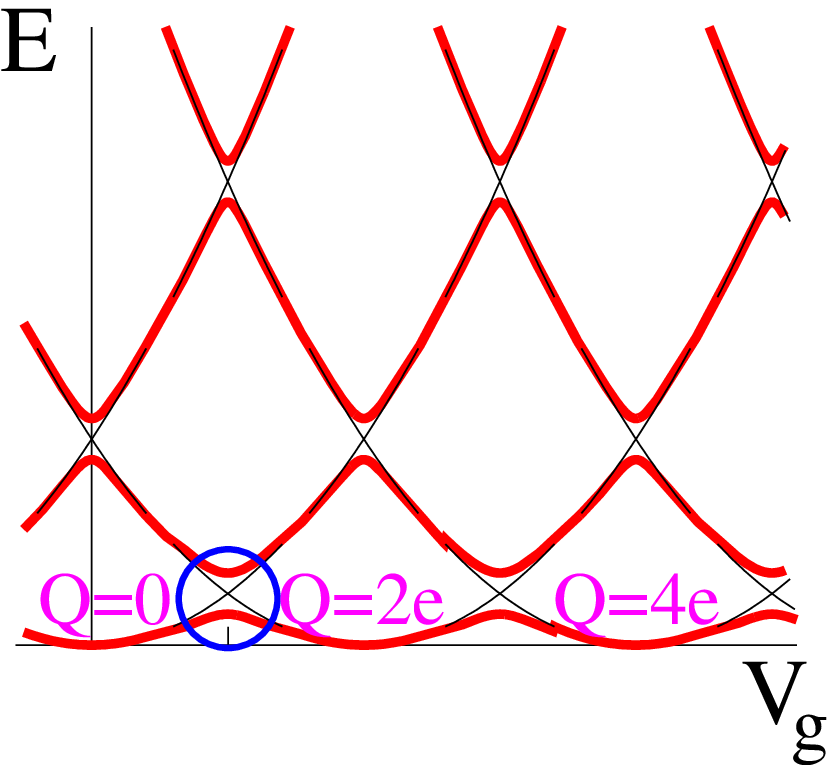}}
\\[-5mm]
{\sf a}\hskip5cm {\sf b}
\caption{a) Josephson charge qubit; b) eigenenergies as functions of the 
gate voltage (for fixed $\Phi_x$)}
\label{Fig:qubit+spectrum}
\end{figure}

When the charging energy dominates, $E_C \gg E_{\rm J}$, 
it is convenient to rewrite Eq.~(\ref{Hbox}) in the eigenbasis of
the number $n$ of Cooper-pair charges \cite{Schoen_Zaikin_Review} 
\begin{equation}
\label{Hbox-charge}
{\cal H} = \sum_{n=-\infty}^\infty \frac{(2ne-Q_{\rm g})^2}{2C}
\ket{n} \bra{n}
-\frac{1}{2}  E_{\rm J}(\Phi_{\rm x})\, \Big( \ket{n}\bra{n+1} + 
\ket{n+1}\bra{n}\Big).
\end{equation}
In this limit it is obvious that eigenenergies form approximately
parabolic bands as a  function of the gate
charge $Q_{\rm g}$, which can be labeled by 
$n$ and are shifted relatively to each other by $2e$ 
(see Fig.~1b). Near the degeneracy points, where two parabolas 
cross, the bands are split by an amount equal to  
$E_{\rm J}(\Phi_{\rm x})$. (A similar picture had been proposed 
in a different context, that of a current-biased low-capacitance 
Josephson junction in 
Refs.~\cite{Widom_BlochOsc_82,Likharev_Zorin_BlochOsc_JLTP85}.) 
If we concentrate on the vicinity of such a degeneracy point the 
system reduces to a quantum two-state system with Hamiltonian
\begin{equation}
\label{H-chargequbit}
{\cal H} = - \frac{1}{2}\Delta E_{\rm ch}(V_{\rm g}) \, \sigma_z-
\frac{1}{2} E_{\rm J}(\Phi_{\rm x}) \, \sigma_x \; .
\end{equation}
We note that by controlling the gate voltage we vary the difference in
the charging energy of the two adjacent charge states, 
$\Delta E_{\rm ch}(V_{\rm g})$, and hence what corresponds to an
effective magnetic field in the $z$-direction. Similarly the
applied flux controls the Josephson coupling energy, corresponding to
a control of the effective magnetic field in the $x$-direction.
In several recent experiments 
\cite{BouchiatPhysScr,Nakamura_Spectroscopy,Nakamura_Nature}
quantum manipulations of Josephson charge qubits
described by (\ref{H-chargequbit}) have been performed.

Also for general relations between the charging energy and the Josephson
coupling energy we have a well-defined quantum-mechanical 
problem. The system has a sequence of levels
$E_i(V_{\rm g},\Phi_{\rm x})$, which depend periodically on the gate 
voltage and the applied flux and can be controlled in this way.
At low temperatures and for
suitable applied ac-fields it may be sufficient to concentrate on the
lowest two of these states \cite{Saclay_Manipulation_Science}. 
The other extreme, $E_{\rm C}\ll E_{\rm J}$, is realized in Josephson flux 
qubits, studied by other
groups \cite{StonyBrook_Cats,Delft_Cats}. Here the control has to be achieved 
solely by applied
fluxes (possibly, different fluxes applied 
to different parts of the circuit).

The control of the qubit by applied voltages and fluxes is always
accompanied by noise. As is clear from Eq.~(\ref{H-chargequbit}) a 
fluctuating control voltage $\delta V(t)$ couples to $\sigma_z$, 
while the flux noise $\delta \Phi(t)$ couples to $\sigma_x$. 
Since these noises are derived from linear circuits they are Gaussian 
and can be modeled by a bath of harmonic oscillators.
They are completely characterized  by their power spectra, which in 
turn depend on the impedance $Z(\omega)$ and the temperature of the 
relevant circuits. 

For instance, the gate voltage fluctuations in the control circuit satisfy
\begin{equation}
S_V(\omega) = 2\,
{\rm Re}\{Z_{\rm t}(\omega)\} \, \hbar \omega \,
\coth\left( \frac{\hbar \omega}{2k_{\rm B}T}\right) \ .
\label{dVdV_Connected}
\end{equation}
They are governed by the total impedance of the circuit
as seen from the qubit, $Z^{-1}_{\rm t}(\omega) \equiv i\omega C_{\rm qb} + 
Z^{-1}(\omega)$, which is determined by the external circuit 
$Z(\omega)$ and the capacitance 
$C_{\rm qb}\equiv(C_{\rm J}^{-1}+C_{\rm g}^{-1})^{-1}$ 
of the qubit in the circuit.
At the degeneracy point
the voltage fluctuations couple linearly to the qubit, via
${\cal H}_1 = -2e(C_{\rm g}/C_{\rm qb})\, \delta V\, \sigma_z$.
In general, this noise is neither longitudinal nor transverse in
the eigenbasis of ${\cal H}_{\rm ctrl}$, rather it is characterized by 
the angle $\theta_V=\tan^{-1}(E_{\rm J}/\Delta E_{\rm ch})$.
At low frequencies the circuit usually behaves as a 
resistor $Z(\omega) = R_V$, and the results for Ohmic dissipation 
discussed in the previous section apply. They are characterized by 
the dimensionless parameter
\begin{equation}
\alpha_V=\frac{4R_V}{R_{\rm K}}
\left(\frac{C_{\rm g}}{C_{\rm qb}}\right)^2
\; ,
\label{gammaV}
\end{equation}
which is determined both by the strength of voltage 
fluctuations of the environment ($\propto R_V$) and by the coupling 
of these fluctuations to the qubit ($\propto C_{\rm g}/C_{\rm qb}$). 
We note that the circuit resistance is compared to the 
(typically much higher) quantum 
resistance $R_{\rm K}=h/e^2 \approx 25.8~{\rm k}\Omega$.
At frequencies considered here the typical impedance of the
voltage circuit  is $R_ V \approx 50~\Omega$. Furthermore, the effect of 
fluctuations is weakened if the gate 
capacitance is chosen small, $C_{\rm g} \ll C_{\rm J}$ . Taking the ratio
$C_{\rm g}/C_{\rm J}=10^{-2}$ one reaches a dissipation as weak as 
$\alpha_V \approx 10^{-6}$, allowing, theoretically, for $\sim 10^6$ 
coherent single-bit manipulations.

Similarly the fluctuations of the external flux can be expressed 
by the effective impedance of the current circuit
 which supplies the flux and the 
mutual inductance characterizing the coupling. Choosing the 
impedance purely real, $R_I$, we have 
\begin{equation}
\label{Eq:alphaI_ch}
\alpha_\Phi=
\frac{R_{\rm K}}{4R_I}
\left(\frac{M}{\Phi_0}
\frac{\partial E_{\rm J}(\Phi_x)}{\partial\Phi_x}
\right)^2
\, , 
\end{equation}
and $\theta_\Phi=\tan^{-1}(\Delta E_{\rm ch}/E_{\rm J})$.
For an estimate we take $R_I\sim 100$~$\Omega$ of the order of the 
vacuum value.  For $M \approx 0.01 - 0.1$~nH and 
$E_{\rm J}^{0}/k_{\rm B} \approx 0.1$~K we obtain $\alpha_\Phi
\approx 10^{-6}  - 10^{-8}$. 

In spite of these favorable estimates for the dephasing rates the actual 
experiments suffer from further, stronger noise sources.
For instance, in charge qubits~\cite{Nakamura_Echo} an important 
contribution comes from ``background charge fluctuations''.  Their origin
may be charge transfer processes between impurities in the substrate,
which typically lead to a $1/f$ power spectrum. This noise source 
can be modeled by further 2-state quantum 
systems~\cite{Paladino_1/f,Nguyen_Girvin_1/f}, however, when many of
them couple weakly to
the TLS they can be approximated again by an oscillator bath 
with appropriate spectrum. We will analyze this model in the next
section. 

There are ways to reduce the effect of the fluctuating environment 
\cite{Saclay_Manipulation_Science}, which can be demonstrated by 
considering the 2-state Hamiltonian (\ref{H-chargequbit}): 
The idea is to tune the gate
voltage to the degeneracy point $V_{\rm g0}$, where the difference in the 
charging energy vanishes, $\Delta E_{\rm ch}(V_{\rm g0}) = 0$,
and, furthermore, to tune the applied flux to a point where
the effective Josephson coupling energy has an extremum,
$\partial E_{\rm J}(\Phi_{\rm x})/\partial 
\Phi_{\rm x} = 0$. Hence the Hamiltonian is
\begin{equation}
\label{H_Saclay}
{\cal H} = - \frac{1}{2} E_{\rm J}(\Phi_{\rm x0}) \, \sigma_x 
-  \frac{1}{2} \frac{\partial \Delta E_{\rm ch}(V_{\rm g})}
{ \partial V_{\rm g}} \, \delta V\, \sigma_z
-   \frac{1}{4} \frac{\partial^2 E_{\rm J}(\Phi_{\rm x})}
{ \partial \Phi_{\rm x}^2} \,\delta\Phi^2  
\, \sigma_x 
 \; .
\end{equation}
At this operating point the energy difference 
between the two relevant states depends only quadratically on 
the fluctuations in either the gate voltage $\delta V(t)$ 
or the applied flux $\delta \Phi(t)$.
On the other hand, one still can perform 
quantum-state manipulations by applying small-amplitude 
ac gate voltages, $V_{\rm ac}(t)$, as is usual in NMR. (Only the noise at 
frequency $\Delta E/\hbar$ couples through this channel; its effect is much 
weaker than that of the pure dephasing if $1/f$ noise is dominant.)
By employing this idea the group in Saclay \cite{Saclay_Manipulation_Science}
has recently observed remarkably long dephasing times of the order
of 1~$\mu$sec. This progress has been one of our motivations to study
noise effects in more general situations.

Another motivation to extend the analysis of the noise comes from the
investigation of quantum measurement devices.  Except when a measurement is
being performed, the detector should be decoupled from the TLS as much as
possible.  Since it is hard to achieve complete decoupling, one usually turns to
zero the linear coupling, but higher-order terms may still be present and affect
the quantum dynamics.  As an example we consider a single-electron transistor
(SET), which can serve as a detector of the quantum state of a Josephson charge
qubit~\cite{Our_RMP,Schoelkopf}.  The interaction between qubit and SET is given
by ${\cal H}_{\rm int} = E_{\rm int}\sigma_z N$, where $N=\sum_i a^{\dag}_{\rm
I,i} a^{\phantom{\dag}}_{\rm I,i}$ is the number of electrons on the central
island of the SET, and $E_{\rm int}$ denotes the capacitive coupling energy.
The electron number $N$ changes due to tunneling to the leads of
the SET.  Thus, in general the SET involves fermionic baths,
but in some cases the problem can be mapped onto a spin-boson
model:  Consider the off-state of the SET with no transport voltage.
In this case tunneling processes  are suppressed at temperatures below
the Coulomb gap of order $E_C$ by the Coulomb blockade, and,
classically, the charge of the central island is fixed at, 
say, $N=0$. On the other hand, virtual higher-order processes
(cotunneling) involve other 
charge states, e.g.  $N=1$, which makes $N$ noisy.  This situation can be mapped
onto (the two lowest states of) a harmonic oscillator, with eigenfrequency
equal to the Coulomb gap $E_C$, coupled to a dissipative bath.  Since the
interaction between the qubit and the oscillator depends on the occupation
number operator of this oscillator, it corresponds to a
nonlinear coupling in the formulation presented above.

In this context we can also mention the 
recent proposal~\cite{INSQUID} of a SQUID-based device which allows 
switching off the linear coupling of a detector to a flux qubit, 
leaving also in this example only 
nonlinear coupling.

These and further examples provide the motivation to study noise effects in 
more general situations than those covered by the usual spin-boson model.
In the next Section we, therefore, investigate Ohmic circuit noise 
as well as $1/f$ noise, 
which couple in a linear or nonlinear way to the two-state quantum system.
The noise can be longitudinal or transverse, which means in the  eigenbasis
of the TLS they couple to $\tau_z$ or $\tau_x$, respectively. 

\section{Extensions of the spin-boson model}
\label{Sec:1/f}

The spin-boson model has been studied
mostly for the specific case where a bath with Ohmic spectrum is coupled
linearly to the spin degree of freedom. One reason is that linear damping,
proportional to the velocity (Ohmic), is encountered frequently in
real systems. Another is that suitable systems with Ohmic
damping show a quantum phase transition at a critical strength 
of the dissipation, with $\alpha_{\rm cr} \sim 1$. 
On the other hand, in the context of quantum-state engineering we 
should concentrate on systems with weak damping,
but allow for general coupling and general spectra of the fluctuations.
In these cases we find qualitatively novel results. 
For instance, a two-level system with sub-Ohmic damping 
(e.g., $1/f$ noise) still shows coherent oscillations,
in spite of a general belief that it should localize in one of its states.
Vise versa, in the super-Ohmic regime, where usually dephasing effects
are believed negligible, we still find that manipulations with sharp pulses
do influence dephasing. Finally we consider nonlinear 
coupling of the noise source to the qubit, which is important 
at symmetry points \cite{Saclay_Manipulation_Science}.    
Here, for definiteness we will consider only one source of fluctuations 
at a time.

\subsection{Sub-Ohmic and $1/f$ noise}
\label{Subsec:_1/f}

As mentioned above several experiments with Josephson circuits 
revealed in the low-frequency range the presence of $1/f$ noise. While the 
origin of this noise may be different in different circuits and requires 
further analysis, it appears that in several cases it derives from
``background charge fluctuations''. In this case  
the noise of the gate charge may be quantified as 
$S_{Q_{\rm g}}(\omega) = \alpha_{1/f}e^2/\omega$. 
Recent experiments~\cite{Nakamura_Echo} yield $\alpha_{1/f} \sim 
10^{-7}$--$10^{-6}$ and indicate that the $1/f$ 
frequency dependence may extend up to high values, of the order of the 
level spacing of the TLS.

One way to model the noise is to use an oscillator bath, where in the
case of  nonequilibrium $1/f$ noise the bath temperature $T_{\rm b}$
should be treated as an adjustable parameter.  
For instance, a bath with $J(\omega) = (\pi/2) \alpha \hbar \omega_s$
gives  at $\omega\ll k_{\rm B}T_{\rm b}/\hbar$  the 
$1/f$ noise spectrum
\begin{equation}
\label{1/f}
S_X(\omega) = \frac{E_{1/f}^2}{|\omega|}
\,.
\end{equation}
with $E_{1/f}^2=2\pi\alpha\hbar\omega_s k_{\rm B}T_{\rm b}$.

Such a model is a particular example
of the so-called {\it sub-Ohmic} damping, defined by spectral densities
\begin{equation}
J_s(\omega) = (\pi/2)\,\hbar \alpha\, \omega_s^{1-s}\omega^s
\qquad\mbox{with } s<1
\,.
\end{equation}
Sub-Ohmic damping was considered earlier, for $0<s<1$,
in the framework of the spin-boson model~\cite{LeggettRMP,WeissBook}, 
but did not attract much attention. It was argued that 
in the presence of  sub-Ohmic dissipation
coherence would be totally suppressed, 
transitions between the states of the
two-level system would be incoherent and take place only at
finite temperatures.  
At zero temperature the system should be localized in one 
of the eigenstates of $\sigma_z$. The localization results from the 
fact that the bath renormalizes the off-diagonal part of the Hamiltonian 
$B_x$ to zero. However, this scenario, while correct for intermediate
to strong damping, is not correct for weak damping. Indeed the NIBA 
approximation~\cite{LeggettRMP}, which was designed to cover
intermediate to strong damping, fails
in the weak-coupling limit for transverse noise. In contrast a more 
sophisticated renormalization procedure~\cite{Kehrein_Mielke_PhysLettA96} 
predicts coherent behavior.
In the context of quantum-state engineering we are interested in 
precisely this {\it coherent sub-Ohmic} regime. We will demonstrate that the 
simple criterion, which was used in Section~\ref{Sec:spin-boson} to distinguish 
between regimes of coherent and incoherent dynamics (i.e., the
comparison of  $\Gamma_\varphi^*$ and $\Delta E$), can be used also
for sub-Ohmic environments. 

We consider the transient coherent dynamics of a spin coupled to a
sub-Ohmic bath. If all the oscillators  
of the bath follow the spin adiabatically, the renormalized matrix element 
$B_x\propto \exp[-\int d\omega J(\omega)/\omega^2]$ is indeed
suppressed, implying an incoherent dynamics. 
For a sub-Ohmic bath the integral diverges at low frequencies. However, a 
finite preparation time $\hbar/\Delta$ (cf.~the discussion in 
Sec.~\ref{Sec:spin-boson}) provides a low-frequency cutoff 
$\Delta$ for the oscillators that 
contribute to the renormalization, thus leading to a finite $B_x$.
The low-frequency oscillators, $\hbar\omega_j\ll\Delta$, contribute to the 
dephasing. 
Their effect is only weakly sensitive to the cutoff $\Delta$, since the 
relevant integrals are dominated by very low frequencies.

For longitudinal noise ($\theta=0$) with
a $1/f$ spectrum one obtains for the 
function $\P(t)$ [Eq.~(\ref{Eq:P(t) factorized})], with logarithmic 
accuracy~\cite{Cottet_Naples}, 
${\rm Re}\,\ln \P(t) \approx - (E_{1/f} t)^2|\ln(\omega_{\rm 
ir}t)|/\pi\hbar^2$, where $\omega_{\rm ir}$ is the (intrinsic) infrared cutoff 
frequency for the $1/f$ noise.\footnote{In an experiment with averaging over  
repetitive measurements it may be determined by the time interval $t_{\rm av}$ 
over which the averaging is performed~\cite{Saclay_Manipulation_Science}. In a 
spin-echo experiment the echo frequency may determine this 
frequency~\cite{Nakamura_Echo}.} From this decay law one
can deduce a dephasing rate, 
\begin{equation}
\Gamma_\varphi^* \approx
\frac{1}{\hbar}\,
E_{1/f}\, \sqrt{\frac{1}{\pi} \ln \frac{E_{1/f}}{\omega_{\rm ir}}} \, .
\label{Gamma_1/f}
\end{equation}
  
Next we consider transverse noise, $\theta = \pi/2$.
As in the Ohmic regime we compare the energy splitting $\Delta E = B_x$ 
to the pure dephasing rate $\Gamma_\varphi^*$ (\ref{Gamma_1/f})
 in order to determine 
whether the Hamiltonian- or noise-dominated regime is realized. 
In the former case, $\Delta E \gg \hbar\Gamma_\varphi^*$, coherent 
oscillations are expected.  This is confirmed by a perturbative analysis, 
similar to the one used to derive 
Eqs.~(\ref{Eq:relaxation},\ref{Eq:dephasing}), which gives
in first order in $\alpha$
\begin{eqnarray}
\Gamma_{\rm relax}&=& \frac{E_{1/f}^2}{\hbar\,\Delta E}
\nonumber
\; ,
\\
\Gamma_\varphi&=&
\frac{1}{2}\;\Gamma_{\rm relax}
\label{Eq:dephasing_1/f}
\; .
\end{eqnarray}

A more detailed analysis reveals that
in the present problem higher-order contributions to the 
dissipative rates may be important as well. The reason is that, 
for $\theta = \pi/2$, the first-order relaxation rate is sensitive to the 
noise $S_X(\omega)$ only at a (high) frequency $\Delta E/\hbar$ where the $1/f$ 
noise is weak, whereas in second order the (diverging) low-frequency part of 
the spectrum becomes important. Employing the diagrammatic technique of 
Ref.~\cite{Schoeller_PRB}, we link the time evolution of the density matrix to
a self-energy, $\Sigma$.
We find that the Laplace transform of the second-order contribution 
$\Sigma^{(2)}(s)$ diverges as $s \rightarrow i\Delta E/\hbar$.
Using first- and second-order terms in $\Sigma$ we solve for the poles
of $\tau_+(s)\propto [s - i \Delta E / \hbar - \Sigma(s)]^{-1}$ to describe the 
short-time behavior and find, with logarithmic accuracy
\begin{equation}
\label{Eq:dephasing_2_order_1/f}
\Gamma_\varphi =
a\;
\frac{E_{1/f}^2}{\hbar\;\Delta E} 
\qquad\mbox{with }
\label{Eq:a}
a \approx
\frac{1}{2\pi}
\ln \frac{E_{1/f}^2}{\hbar\,\omega_{\rm ir}\;\Delta E}
\,. 
\end{equation}
Thus, we see that the second-order correction to the self-energy
changes the result  
for the dephasing rate considerably. As a consequence, for 
transverse noise the ratio $\Gamma_\varphi/\Gamma_{\rm relax}$, which
in first order takes the value $1/2$,
is shifted substantially towards higher values. It is interesting that in the 
recent 
experiments of Ref.~\cite{Saclay_Manipulation_Science} a ratio
$\Gamma_\varphi/\Gamma_{\rm relax}\approx 3$ was reported.
A more detailed description of the second order calculation 
will be presented elsewhere. Finally we note that 
Eq.~(\ref{Eq:dephasing_2_order_1/f}) indeed confirms 
the assumption of underdamped coherent oscillations, as 
$\hbar \Gamma_{\rm relax},\hbar \Gamma_\varphi \ll \Delta E$ in the 
Hamiltonian-dominated regime.  

Notice that for a sub-Ohmic bath with $-1<s<1$
due to a high density of low-frequency oscillators the dephasing persists at 
$T_{\rm b}=0$: $|\langle 
\tau_+(t) \rangle| \propto \exp[-\alpha (\omega_st)^{1-s}/(s+1)]$ from
which we read off the dephasing rate 
$\Gamma_\varphi^* \propto [\alpha/(s+1)]^{1/(1-s)}\omega_s$ 
(cf. Ref.~\cite{Unruh}).

In the noise-dominated regime,  $\Delta E \ll \hbar\Gamma_\varphi^*$, 
the dynamics is incoherent as we know from earlier 
work~\cite{LeggettRMP,WeissBook}.
Thus, our criterion for the coherent regime,
$\Delta E \gg \hbar\Gamma_\varphi^*$,
is valid also for $1/f$ and 
sub-Ohmic environments. For $0<s<1$, the critical damping strength
at which $\Delta E \sim \hbar\Gamma_\varphi^*$ coincides with that obtained in 
Ref.~\cite{Kehrein_Mielke_PhysLettA96} and with the 
boundary of the applicability range of  NIBA~\cite{LeggettRMP}.

\subsection{Super-Ohmic case}

Next we discuss a super-Ohmic bath with, e.g., $s=2$ and
$J_s(\omega) = (\pi/2)\,\hbar \alpha\, \omega^2/\omega_s$.
At zero temperature this leads to 
\begin{equation}
{\rm Re}\;K(t) = \frac{2\alpha}{\omega_s}\left(\frac{\sin(\omega_c t)}{t} 
-\omega_c \right) \, .
\end{equation} 
Thus, within a short time of order $\omega_c^{-1}$
the exponent ${\rm Re}\;K(t)$ saturates at a finite value 
${\rm Re}\;K(\infty) = - 2\alpha\omega_c/\omega_s$. While usually 
this reduction of the off-diagonal elements of the density matrix
is not denoted as dephasing (since coherent oscillations persist), 
from the point of view of quantum-state engineering it is a 
relevant loss of phase coherence, which introduces errors.
We also note that this reduction strongly depends on 
the frequency cutoff $\omega_c$. Recalling that this cutoff is determined 
by the quantum-state manipulation (e.g., the width of a $\pi/2$-pulse) we 
observe that in the super-Ohmic regime
the (weak) dephasing is strongly sensitive to the details of the preparation 
procedure.  
 
\subsection{Nonlinear coupling to the noise source}

As motivated in the preceding section we study next a generalization 
of the spin-boson model  to include nonlinear coupling of the noise
source to  the variables of the TLS. 
The Hamiltonian in the eigenbasis of the unperturbed system 
(with Pauli matrices denoted by $\tau$) thus reads
\begin{equation}
\label{Eq:Spin_Boson_Nonlinear}
{\cal H} = -{1\over 2}\Delta E \tau_z  
+ (\sin\theta\;\tau_x + \cos\theta\;\tau_z) 
\;\frac{X^2}{E_0}
+ {\cal H_{\rm b}}
\ . 
\end{equation}  
Since the  bath ``force'' operator
$X=\sum_j c_j (a^{\phantom \dagger}_j+a^\dagger_j)$  has dimensions 
of energy, for nonlinear coupling a new energy scale, $E_0$, 
appears. In specific systems this energy scale is a characteristic 
scale of the system. E.g., for a Josephson charge qubit coupled 
quadratically to the flux noise [Eq.~(\ref{H_Saclay})] we have 
$E_0 = E_{\rm J}$. 

When considering a quadratic coupling we need to know the 
statistical properties of $X^2(t)$. While $X$ is Gaussian-distributed,
$X^2$ is not. However, for a first-order perturbative 
analysis this fact is irrelevant and we can assume that $X^2$ is Gaussian as 
well, with a 
width that is fixed by the correlator
$S_{X^2}(\omega) \equiv \left\langle \{ X^2(t), X^2(t') \} 
\right\rangle_\omega$. For baths with spectral densities which are
regular at low frequencies it can be shown that this approximation is
sufficient to describe the initial stage  
of the dephasing process for $\Gamma_\varphi t\ll1$, and 
it provides a reliable estimate for the dephasing time. 
For singular spectra 
(e.g., for $1/f$ noise) a more detailed analysis is needed, and non-Gaussian 
corrections may appear. 
Note that in order to evaluate the symmetrized correlator
$S_{X^2}(\omega)$ in general one needs to know the full correlator  
$\langle X(t)X(t') \rangle$ and not only the symmetrized one 
$S_X(\omega)$. For an Ohmic spectrum (\ref{Eq:Linear_Spectrum})
the symmetrized and antisymmetrized spectra are related by detailed balance,
 and  we find
\begin{equation}
S_{X^2}(\omega) \sim
\alpha^2 \hbar \max\{(\hbar\omega)^3,(k_{\rm B}T)^3\}
\,.
\label{Eq:OhmicX2}
\end{equation} 
For $1/f$ noise (\ref{1/f}) (assuming that the 
antisymmetric part diverges weaker than $1/\omega$ and can be
neglected)  we find
\begin{equation}
S_{X^2}(\omega) = \frac{2}{\pi}
\frac{E_{1/f}^4}{\omega}
\ln
\frac{\omega}{\omega_{\rm ir}}
\,.
\label{Eq:1/fX2}
\end{equation}

Consider first the longitudinal coupling ($\theta = 0$). For 
an Ohmic bath at low frequencies we have $S_{X^2}(\omega=0) \sim 
\alpha^2 \hbar (k_{\rm B}T)^3$. Therefore,
the pure dephasing rate is given by
\begin{equation} 
\Gamma_\varphi^* = \frac{S_{X^2}(\omega=0)}{\hbar^2 E_0^2} 
\sim
\frac{\alpha^2 (k_{\rm B}T)^3}{\hbar E_0^2}
\ .    
\end{equation}

For $1/f$ noise, the correlator $S_{X^2}(\omega)$ also exhibits a
$1/\omega$ divergence. Hence  
in analogy to the linear-coupling case the dephasing is governed by 
${\rm Re}\ln\P(t) = - [ E_{1/f}^2 t \ln(\omega_{\rm ir}t) /\pi\hbar E_0]^2
$,
i.e., the characteristic dephasing rate is   
\begin{equation} 
\Gamma_\varphi^* = \frac{E_{1/f}^2}{\pi\,\hbar\,E_0}
\;\ln
\frac{E_{1/f}^2}{\omega_{\rm ir}E_0}
\,.    
\end{equation}

In the transverse coupling case ($\theta = \pi/2$) 
the relaxation and dephasing rates are again given by the noise 
power spectrum $S_{X^2}$ at frequency $\Delta E/\hbar$.
In the lowest order we have $\Gamma_{\rm relax}=S_{X^2}(\Delta E/\hbar)/E_0^2$ 
for both Ohmic and $1/f$ noise sources.
In the case of $1/f$ noise we should again take into account second-order 
corrections to the self-energy since they probe the divergent low-frequency part 
of the noise power spectrum. In this way we get $\Gamma_\varphi =
b\;\Gamma_{\rm relax}$ with  $b\sim 1$. 

\subsection{Summary of results}

In the table we summarize our results for the relaxation and pure dephasing
rates, $\Gamma_{\rm relax}$ and $\Gamma_\varphi^*$,
of a two-state quantum system subject to the different
sources of noise, with coupling which is longitudinal or transverse to the qubit
(as compared to its eigenbasis), linear or quadratic, with an Ohmic
(\ref{Eq:Linear_Spectrum}), (\ref{Eq:OhmicX2}) 
or a $1/f$ spectrum (\ref{1/f}), (\ref{Eq:1/fX2}).
For the sake of 
brevity we put $\hbar=k_{\rm B}=1$ in the table and omit factors of order one in 
some expressions. Complete results 
can be found in the text.

\vspace{5mm}
\begin{tabular}{c|l|l}
&
{\bf longitudinal} ($\parallel$)
&
{\bf transverse} ($\perp$)
\\
&
($\Rightarrow$ pure dephasing)
&
($\Rightarrow$ relaxation + dephasing)
\\
\multicolumn{2}{l}{{\bf linear coupling:}}\\[3mm]
&$\displaystyle{\cal H} = \frac{1}{2}\Delta E\;\rho_z
+ X\;\rho_z$
&
$\displaystyle{\cal H} = \frac{1}{2}\Delta E\;\rho_z
+ X\;\rho_x$
\\
\hline
&&\\
``Ohmic''
&
$\Gamma_\varphi^* = S_X(\omega =0) \sim \alpha T$ 
&
$\Gamma_{\rm relax} = S_X(\Delta E)$
\\
&&
$\Gamma_\varphi=\Gamma_{\rm relax} /2 $
\\
\hline
&&\\
1/f
&
$\displaystyle
\langle\tau_+(t)\rangle\sim e^{\displaystyle -
E_{1/f}^2\, t^2\;\ln t}
$
&
$\Gamma_{\rm relax} = S_X(\Delta E)$
\\
&
$\displaystyle\Gamma_\varphi^* = E_{1/f}
\;\ln^{1/2}(E_{1/f}/\omega_{\rm ir})$
&
$\Gamma_\varphi=a\;\Gamma_{\rm relax}$\ \ ,\ $a\sim 1$
\\
\\

\multicolumn{2}{l}{{\bf 
quadratic coupling:}}\\
&&\\
&$\displaystyle{\cal H} = \frac{1}{2}\Delta E\;\rho_z
+ \frac{1}{E_0} X^2\;\rho_z$
&
$\displaystyle{\cal H} = \frac{1}{2}\Delta E\;\rho_z
+ \frac{1}{E_0} X^2\;\rho_x$
\\
\hline
&&\\
``Ohmic''
&
$\Gamma_\varphi^* = \displaystyle\frac{1}{E_0^2} S_{X^2}(\omega
=0) \sim \alpha^2 \frac{T^3}{E_0^2}$
&
$\Gamma_{\rm relax} = \displaystyle\frac{1}{E_0^2} S_{X^2}(\Delta E)$
\\
&&
$\Gamma_\varphi=\Gamma_{\rm relax} /2 $
\\
\hline
&&\\
1/f
&
$\displaystyle
\langle\tau_+(t)\rangle\sim e^{\displaystyle
-(E_{1/f}^4/E_0^2)\;t^2 \,\ln^2 t}
$
&
$\Gamma_{\rm relax} =\displaystyle \frac{1}{E_0^2}
\; S_{X^2}(\Delta E)$
\\
& $\displaystyle
\Gamma_\varphi^* =
\frac{E_{1/f}^2}{E_0}
\;\ln(E_{1/f}^2/E_0\,\omega_{\rm ir})
$
&
$\Gamma_\varphi= b\;\Gamma_{\rm relax}$\ \ ,\ $b\sim 1$
\end{tabular}

\section{Summary}

To summarize, motivated by recent experiments on Josephson-jucntion circuits, we 
have considered dephasing effects in a two-level system due to noise sources 
with various spectra  (incl.\ Ohmic and $1/f$), coupled in a linear or 
nonlinear, longitudinal or transverse way to the TLS.

We have shown that the dephasing can be sensitive to the details of
the initial-state preparation. For instance, a finite preparation time 
$\sim\hbar/\Delta$ introduces an upper cutoff $\Delta$ on the frequency of 
environmental modes that contribute to dephasing, whereas the higher-frequency 
modes merely renormalize the parameters of the TLS. In particular, for an Ohmic 
environment we find a power-law dephasing at $T=0$, sensitive to this cutoff 
frequency.

We have also linked the mentioned renormalization effects to the behavior
of a response funtion of the TLS, which exhibits features known for the 
orthogonality catastrophe, including a power-law divergence above a threshold.

Noise with a $1/f$ spectrum we modeled by a sub-Ohmic bath and found a 
simple criterion for coherent behavior in
sub-Ohmic environments. For transverse $1/f$ noise we found that, due to 
infrared divergences, one needs to take into account second-order contributions 
in a perturbative analysis. This can increase substantially the ratio between 
the relaxation and dephasing times, which takes the value $1/2$ in first-order 
calculations. These findings may contribute to the understanding of the 
experimental results of Ref.~\cite{Saclay_Manipulation_Science}, where a higher 
ratio ($\sim3$) was found.

We have also analyzed the dynamics in the case of an environment with Ohmic or 
$1/f$ noise power, which is coupled nonlinearly to the TLS.
In the Ohmic case (e.g., for a single-electron transistor coupled to a 
Cooper-pair box) the dephasing rate scales as $T^3$, and by biasing the SET at a 
special point it can be further suppressed to a $T^5$-dependence. This 
should be contrasted, e.g., with a quantum point contact which, even in the 
off-state, couples linearly to the TLS and influences it strongly, 
$\Gamma_\varphi\propto T$.
We reduced the description of a TLS coupled nonlinearly to a noise source with 
$1/f$ spectrum to that of a TLS with linear coupling to an effective $1/f$ noise 
source. The results for the relaxation and dephasing times in this situation are 
relevant for recent experiments~\cite{Saclay_Manipulation_Science}.   

\section{Acknowledgements}

We thank M.H.~Devoret, D.~Esteve, Y.~Gefen, D.~Golubev, Y.~Imry, S.K.~Kehrein, 
A.D.~Mirlin,  Y.~Nakamura, A.~Rosch, D.~Vion, and A.D.~Zaikin for stimulating 
discussions. 
The work is part of the EU IST Project SQUBIT
and of the {\bf CFN} (Center for Functional Nanostructures) 
which is supported by the DFG (German Science Foundation).
Y.M. was supported by the Humboldt Foundation, the BMBF and the ZIP programme of 
the German government.


\end{document}